\documentclass[12pt,a4paper]{article}

\usepackage{graphicx}
\usepackage{bm}

\newcommand{\be}{\begin{equation}}
\newcommand{\ee}{\end{equation}}
\newcommand{\bel}[1]{\be\label{#1}}
\newcommand{\re}[1]{Eq.~(\ref{#1})}

\newcommand{\ds}{\displaystyle}
\newcommand{\ptt}{\partial_t}
\newcommand{\ptx}{\partial_x}
\newcommand{\hsp}{\hspace*{1pt}}
\begin{document}
\renewcommand{\thefootnote}{\arabic{footnote}}

\begin{center}
{\Large\bf Influence of irradiation on the space--time structure
of~shock waves}\\[5mm]
{\bf J.A. Maruhn$^{\,1}$, I.N.~Mishustin$^{\,1,2,3}$,
L.M.~Satarov$^{\,1,2}$}
\end{center}
\begin{tabbing}
\hspace*{1.5cm}\=
${}^1$\={\it Institut~f\"{u}r~Theoretische~Physik,
J.W.~Goethe~Universit\"{a}t,}\\
\>\>{\it D--60054~Frankfurt~am~Main,~\mbox{Germany}}\\
\>${}^2$\>{\it The Kurchatov~Institute, Russian Research Centre,}\\
\>\>{\it 123182~Moscow,~\mbox{Russia}}\\
\>${}^3$\>{\it The Niels~Bohr~Institute,
DK--2100~Copenhagen {\O},~\mbox{Denmark}}\\
\end{tabbing}

\begin{abstract}
The long--range energy deposition by heavy--ion beams makes
new types of shock wave experiments possible in the
laboratory. We have investigated a situation that is of
relevance to supernova dynamics in astrophysics, where
a shock wave is irradiated by a flux of neutrinos
depositing energy throughout the shock wave and surrounding
matter, thus changing the behaviour of the running shock.
We have carried out fluid--dynamical simulations to study
generic features of stimulated shock waves. First we
consider an idealized case assuming uniform energy
deposition into a planar shock wave propagating through
an ideal gas. Then we investigate more realistic
situations realizable in laboratory experiments with heavy--ion beams.
We have found that energy deposition leads to two important effects:
acceleration of the shock front and decay of the shock strength.
The possibility of laboratory experiments is briefly discussed.

\end{abstract}

\baselineskip 24pt

\section{Introduction}

Supernova explosions represent one of the most spectacular phenomena
in our Universe. In recent years much effort has been devoted to
developing realistic theoretical models of this complex process (see
the recent review~\cite{Lei01}).  The main mechanism includes the
shock wave generation upon the bounce of the infalling iron core of a
massive star, but detailed calculations showed that the prompt bounce
shock mechanism does not lead directly to ejection of the stellar
envelope. Due to severe energy losses due to, e.g.,
photodisintegration of iron nuclei, the outward shock wave stops
inside the iron core. Recently several new mechanisms of the shock
revival have been suggested~\cite{Bet85,Jan01} which include neutrino
heating and convection in the postshock matter. These processes raise
the postshock pressure and provide additional energy for the shock
wave expansion. The calculations performed by various groups differ,
however, in conclusions concerning the possibility of successful
shocks.  In this rather unclear situation a better understanding of
the stimulated shock dynamics is highly desirable~\cite{Jan01}.
Fortunately, intense heavy--ion beams open a new possibility to study
such processes in laboratory experiments.

In the first part of this paper we formulate a simple model to study
the evolution of a planar shock wave under influence of homogeneous
irradiation. We have performed detailed fluid--dynamical calculations
for the ideal gas equation of state. The matter flow behind the shock
front is rather complicated, resembling a decay of an initial
discontinuity~\cite{Lan87}. The calculations show that under
irradiation the shock front accelerates, but the density jump at this
front diminishes. It is remarkable that a self--similar regime of
hydrodynamic flow is established at late times.

In the second part of the paper we generalize our model to consider
more realistic situations which may be easily reproduced in
laboratory conditions. Namely, we consider the case when the initial
shock wave is created by a beam with finite extension in the
transverse directions. Specifically, we study the scenario in
which the first beam creates a shock wave in a cylindrical target and
then, when it has developed sufficiently, the target is irradiated
by a second beam.  The target length is chosen in such a way that the
shock generated at the initial stage is completely within the
deposition region of the second beam%
\footnote{%
 It is, of course,
 irrelevant how the shock itself is created --- instead of a heavy-ion
 beam a laser could be used, for example, --- and the second beam
 could come from any suitable direction.
}.
In our simulations we assume that both are heavy--ion beams with equal
properties and are directed along the target axis from opposite
directions.  General properties of hydrodynamical flow in such targets
have been studied earlier in Ref. \cite{Mar98}. Another interesting
phenomenon is predicted for the case of a single beam irradiation with
constant temporal profile.  It is shown that with increasing
irradiation time, when radial flow of matter behind the shock front
becomes noticeable, the shock wave in its central parts decays as
compared to initial stages. The origin of this phenomenon consists in
the increased range of the bombarding particles at later stages of
target irradiation.

The article is organized as follows: in Sect.~II a simple
fluid--dynamical model is formulated for the case of homogeneous
energy deposition. Then in Sect.~III this model is used to study the
dynamics of a planar shock wave in matter with an ideal gas equation
of state. The asymptotic regime of the matter flow is considered by a
semi--analytic method outlined in the Appendix. The results of more
realistic calculations and suggestions for future experiments with
heavy--ion beams are given in Sect.~IV.  The main results of the present
paper are summarized in Sect.~V.

\section{Equations of fluid dynamics}

Let us consider a beam of energetic particles
irradiating a target with the mass
density~$\rho$\,. Often we shall speak about photon
irradiation, but the same approach can be applied for other
penetrating particles, e.g. for neutrino beams.
The strength of irradiation is characterized by the energy
deposited per  unit space--time volume in the
local rest frame of the target,
\bel{depr}
\frac{\ds dE}{\ds dt\,d^3r}=\lambda\rho,
\ee
where $\lambda$ is the specific deposition rate.
Let the target consist of atoms of different kinds
$i$ with particle number densities $n_i$\,. In the case of
monochromatic photons with the energy flux $I(\omega)$ one has
\bel{depe}
\lambda\rho=I(\omega)\sum_i {n_i<\sigma_{i}^{\mathrm{abs}}(\omega)>}\,,
\ee
where $\hbar\omega$ is the photon energy,
$<\sigma_i^{\mathrm{abs}}>$ is the cross section
of photon absorption on atoms of the $i$--th kind. The angular brackets
denote averaging over their momentum distribution.
In this paper we neglect internal heat transport processes and
assume $\lambda$ to be constant in time and space.
The last assumption implies that the external
radiation is homogeneous,  Doppler effects
are not important and the photon absorption length,
\mbox{$(\sum\limits_i n_i <\sigma_{i}^{\mathrm{abs}}>)^{-1}$}, is large
as compared to the characteristic size of a target.

Assuming that energy deposition proceeds under condition of local
thermodynamic equilibrium, we study
the dynamics of target flow by solving nonrelativistic fluid--dynamical
equations. Let us consider first a one--dimensional
case when matter moves along the $x$ axis
with the velocity $v\equiv v_x$\,. The equations of motion then take
the form
\begin{eqnarray}
&&\ptt\hsp\rho+\ptx\hsp(\rho v)=0\,,\label{fl21}\\
&&\ptt\hsp(\rho v)+\ptx\left(P+\rho v^2\right)=0\,,
\label{fl22}\\
&&\ptt\left(\epsilon+\rho v^2/2\right)+
\ptx\left[v\left(\epsilon+P+\rho v^2/2\right)\right]
=\lambda\rho\,.\label{fl23}
\end{eqnarray}
Here $P$ is pressure and $\epsilon$ is the energy density.
By using Eqs.~(\ref{fl21})--(\ref{fl22}) and thermodynamic relations
one may rewrite \re{fl23} in the equivalent form:
\bel{fl23a}
T\left(\ptt+v\hsp\ptx\right) s_m=\lambda\,,
\ee
where $T$ is temperature and $s_m$ is the specific entropy.
This equation shows that $\lambda$ is, in fact, the specific
rate of heat deposition into the target matter.

In the following we assume that the target matter can be
regarded as an ideal Boltzmann gas with one species
of molecules and a constant ratio of heat capacities
\mbox{$\gamma=C_p/C_{\mathrm{v}}$}. The equation of state of such matter
has a very simple form
\bel{eos1}
P=(\gamma-1)\,\epsilon=\rho RT/M\,,
\ee
where $R$ is the gas constant and $M$ is the molar mass.
The specific entropy in this case is
\bel{eos2}
s_m=\frac{\ds R}{\ds (\gamma-1) M}\ln{\frac{\ds P}
{\ds\rho^\gamma}}+{\mathrm{const}}\,.
\ee

Using \re{eos1} and introducing the specific volume \mbox{$V=1/\rho$}
one can rewrite \mbox{Eqs.~(\ref{fl21})--(\ref{fl23})} as
follows
\begin{eqnarray}
&&(\ptt+v\hsp\ptx)\hsp V=V\ptx v\,,\label{fl31}\\
&&(\ptt+v\hsp\ptx)\hsp v=-V\ptx P\,,\label{fl32}\\
&&V (\ptt+v\hsp\ptx)\hsp P+\gamma P\hsp(\ptt+v\hsp\ptx)\hsp
V=(\gamma-1)\,\lambda\,.\label{fl33}
\end{eqnarray}

It is convenient to introduce instead of $P, V, v$ new
dimensionless quantities \mbox{$\bar{P}=P/P_0$},
\mbox{$\bar{V}=\rho_0 V$}, \mbox{$\bar{v}=v/v_0$}
where $P_0$ and $\rho_0$ are
initial pressure and density, and
\bel{vesc}
v_0=\sqrt{P_0/\rho_0}\,.
\ee
Let us make the transition from $t, x$ to
dimensionless variables $\bar{t}=t/t_0$ and $\bar{x}=x/x_0$\,,
where
\bel{txsc}
t_0=v_0^2/\lambda\,,\,\,x_0=v_0 t_0\,.
\ee
Now\, Eqs.~(\ref{fl31})--(\ref{fl33})\, may be rewritten
in the same form, but\, with the repla\-ce\-ments
\mbox{$P,V,v,t,x\to\bar{P},\bar{V},\bar{v},\bar{t},\bar{x}$}
and \mbox{$\lambda\to 1$}\,. If the initial conditions
do not impose additional scales, there is no need in
solving fluid--dynamical equations for different deposition
rates~$\lambda$\,.
It is sufficient to find the solution only for one fixed
value of $\lambda$\,, then the above scaling can be used
to obtain the solution for other values.
For example, this scaling can be applied
for the shock wave initial conditions (see the next section).

We close the general discussion by defining
three families of characteristics
$x\raisebox{-.8ex}{$\scriptstyle C_+$},\,
x\raisebox{-.8ex}{$\scriptstyle C_-$},\,
x\raisebox{-.8ex}{$\scriptstyle C_0$}$\,.
They are  solutions of the following differential equations
\begin{eqnarray}
\dot{x}(t)&=&v+c_s\hspace*{1.5cm}(C_+)\,,\label{char11}\\
\dot{x}(t)&=&v-c_s\hspace*{1.5cm}(C_-)\,,\label{char12}\\
\dot{x}(t)&=&v\hspace*{2.25cm}(C_0)\,,\label{char13}
\end{eqnarray}
where $v$ and $c_s$ are taken at $x=x(t)$\,.
As well--known~\cite{Lan87}, these characteristics describe
propagation of small disturbances of fluid--dynamical
quantities. In particular, the entropy disturbances
propagate along the $C_0$ characteristics (the latter are
also the collective flow trajectories).

In the following we assume that irradiation starts at
$t=0$ and the initial profiles $\rho (x,0)$\,, $v (x,0)$
and $P(x,0)$ are known function of $x$\,.
Let us consider first the case of homogeneous initial
conditions, when \mbox{$\rho (x,0)=\rho_0$},\,
\mbox{$v(x,0)={\mathrm{v}}_0$} and \mbox{$P(x,0)=P_0$}.
In this case the solution of Eqs.~(\ref{fl31})--(\ref{fl33})
is trivial:
\begin{eqnarray}
V(x,t)&=&1/\rho_0\,,\label{sf11}\\
v(x,t)&=&{\mathrm{v}}_0\,,\label{sf12}\\
P(x,t)&=&P_0+(\gamma-1)\hsp\rho_0\hsp\lambda\hsp t\,.\label{sf13}
\end{eqnarray}
In the same case the adiabatic sound velocity is equal to
\bel{adsv}
c_s=\sqrt{\gamma PV}=
\sqrt{c^2_0+\gamma\hsp (\gamma-1)\lambda\hsp t}\,,
\ee
where $c_0=\sqrt{\gamma P_0/\rho_0}$ is the initial sound
velocity.

For homogeneous initial conditions, substituting
(\ref{sf12}), (\ref{adsv}) into Eqs.~(\ref{char11})--(\ref{char13}),
one can find characteristics analytically:
\begin{eqnarray}
x\raisebox{-.8ex}{$\scriptstyle C_\pm$}(t)&=&{\mathrm{v}_0}t\pm\frac{2}
{3\hsp\gamma\hsp(\gamma-1)\hsp\lambda}
\left[c^2_0+\gamma\hsp (\gamma-1)\lambda\hsp t\,\right]^{3/2}
+{\mathrm{const}}\,,\label{char21}\\
x\raisebox{-.8ex}{$\scriptstyle C_0$}(t)&=&{\mathrm{v}_0}t+
{\mathrm{const}}\,.\label{char22}
\end{eqnarray}
According to \re{char21}, at large irradiation times the
$C_\pm$ characteristics approach asymptotically the
same lines which do not depend on initial conditions:
\bel{asch}
x\raisebox{-.8ex}{$\scriptstyle C_\pm$}(t)
\simeq\pm\,\xi_0\sqrt{\lambda t^3}
\hspace*{1cm}(t\gg c_0^2/\lambda\,, {\mathrm{v}}_0^2/\lambda)\,,
\ee
where
\bel{xi0d}
\hspace*{-1.6cm}\xi_0=\frac{2}{3}\sqrt{\gamma\hsp(\gamma-1)}\,.
\ee
\section{One--dimensional shock waves}

\subsection{Numerical results}

Let us consider a one--dimensional shock wave
propagating to the left along the $x$~axis (directed to the
right) and denote the position of its front by $x_{\mathrm{sh}}(t)$\,.
For brevity we introduce the vector $\bm{y}=(V,v,P)$
which combines the set of fluid--dynamic quantities
in a compact way.
It is assumed that at $x=x_{\mathrm{sh}}$ this vector
jumps from \mbox{$\bm{y}_1\equiv\bm{y}(x_{\mathrm{sh}}-0,t)$} to
\mbox{$\bm{y}_2\equiv\bm{y}(x_{\mathrm{sh}}+0,t)$}. In the
rest--frame of the shock wave (moving with the velocity
$D=|\dot{x}_{\mathrm{sh}}|$ with respect to the laboratory
frame) the fluxes of mass, energy, and momentum
should be continuous at $x=x_{\mathrm{sh}}$\,. This gives three
relations~\cite{Lan87} connecting $D$ and the components of
vectors $\bm{y}_{1,2}$\,. In the case of an ideal gas one obtains%
\footnote{%
 It is worth noting that the last equality in \re{shwr} is in fact the
 Rankine--Hugoniot relation. Excluding $D$ from the first two relations
 one can find that the velocity jump\label{fot1} is equal to
 \mbox{$v_1-v_2=\sqrt{(P_2-P_1)\hsp (V_1-V_2)}$}.
}
\bel{shwr}
\frac{v_1+D}{V_1}=\frac{v_2+D}{V_2}=
\sqrt{\frac{P_2-P_1}{V_1-V_2}}=
\sqrt{\frac{\gamma\,(P_1+P_2)}{\,V_1+V_2}}\,,
\ee
where $v_{1,2}$ are defined in the laboratory frame.
Therefore, at given $\bm{y}_1$, all characteristics
of the shock wave can be determined if one of the
quantities \mbox{$D,\,P_2,\,V_2,\,v_2$} is known.
One should bear in mind that the relations (\ref{shwr})
are based only on the local conservation laws
and hold also for shock waves with energy deposition.
However, as will be shown below, in this situation
the characteristics of the shock wave (in particular, $D$)
are in general time--dependent.

It is further assumed that a step--like one--dimensional
shock wave was created in a target at $t<0$\,. If pressure
and density before and behind the shock front do not change with
$x$ and $t$\,, the shock velocity $D$ will be constant.
Below we study the dynamics of this ``initial'' shock wave
after switching on the energy deposition at $t=0$\,.
Eqs.~(\ref{fl31})--(\ref{fl33}) are solved numerically
by using the FCT--algorithm~\cite{Bor73}.
The calculations are performed in the rest frame of the
initial shock. Choosing the origin of the $x$ axis at the
position of the shock front, one has
$x_{\mathrm{sh}}(t)=0$ at $t<0$\,. To be more specific, the
following initial conditions are applied
\bel{inc1}
\bm{y}(x,0)=\bm{y}_{10}\Theta (-x)+\bm{y}_{20}\Theta (x)\,,
\ee
where $\Theta (x)\equiv \frac{1}{2}\hsp (1+{\mathrm{sign}}\,x)$\,,
and $\bm{y}_{10}$ and $\bm{y}_{20}$ denote the fluid--dynamical
quantities in front and behind of the initial shock front,
respectively.
We choose these quantities by fixing the density
($\rho_0\equiv\rho_{10}=1/V_{10})$ and pressure
($P_0\equiv P_{10}$) at $x<0$ as well as the pressure~($P_{20}$)
at~$x>0$\,.

Using the relations (\ref{shwr}) one has
\begin{eqnarray}
V_{20}&=&\frac{(\gamma-1)\hsp\chi+\gamma+1}
{(\gamma+1)\hsp\chi+\gamma-1\hsp}\,V_{10}\,,\label{inc21}\\
v_{10}&=&v_{20}\frac{V_{10}}{V_{20}}=D(0)=
v_0\sqrt{\frac{(\gamma+1)\hsp\chi+\gamma-1}{2}},\label{inc22}
\end{eqnarray}
where $\chi=P_{20}/P_{10}$ is the initial pressure ratio,
$v_0$ is defined by \re{vesc} and the velocities
\mbox{$v_{i0}\,\,(i=1,2)$} are taken in the rest frame
of the shock front.

Most calculations have been performed for $\gamma=5/3$ and
$\chi=4$\,.
In this case the initial density jumps at the shock front from
$\rho_0$ to $2.125\,\rho_0$ and the flow velocity jumps from
\mbox{$v_{10}\simeq 2.380\,v_0$} to
\mbox{$v_{20}\simeq 1.120\,v_0$}.
Some results of numerical calculations for these initial
conditions are shown in Figs.~1--4. The results are
presented using the scaled variables $t/t_0$ and
$x/x_0$\,, with $t_0,\,x_0$ defined in \re{txsc}. As explained in
the preceding section, in such a representation
the whole $\lambda$ dependence is contained only in the scales $t_0$
and $x_0$.

\begin{figure}[b]
\vspace*{-2cm}
\includegraphics[width=15cm]{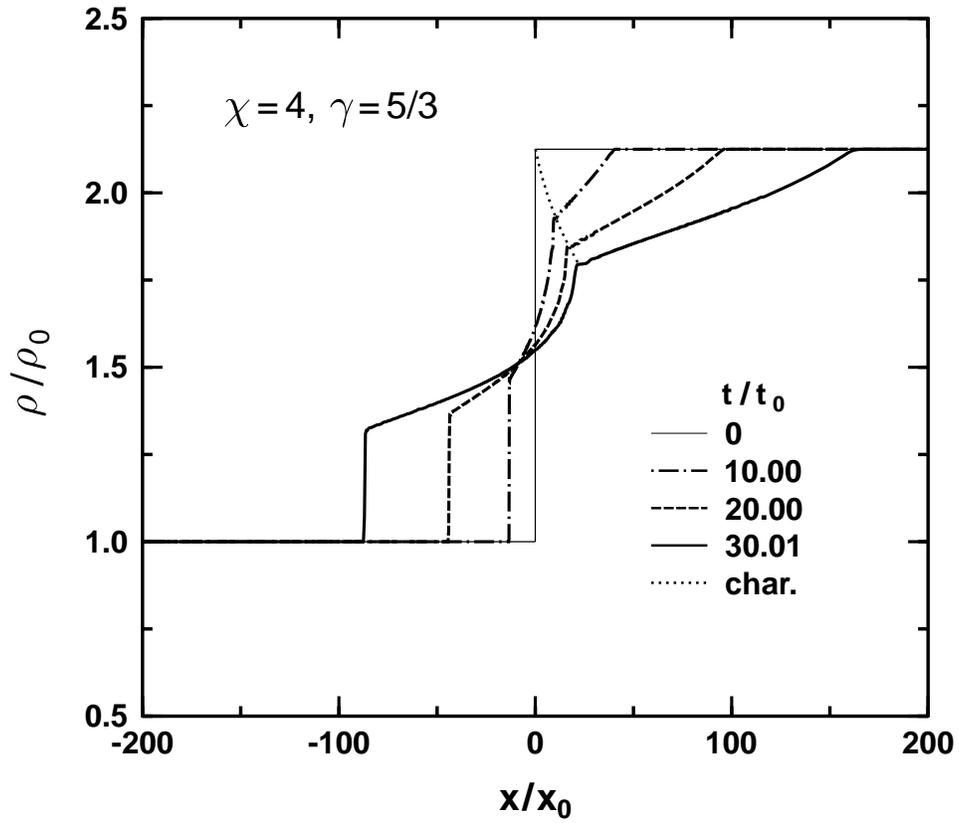}
\vspace*{-4.5cm}
\caption
{Density profiles for different times of shock wave irradiation.
The initial pressure ratio $\chi=4$\,.
Scales $x_0$ and $t_0$ are given by
Eqs.~(\ref{vesc})--(\ref{txsc}). The dotted curve shows
densities on the $C_0$ characteristic with \mbox{$x(0)=0$}.}
\label{figure1}
\end{figure}

Fig.~\ref{figure1} shows the density profiles at different times from the
beginning of irradiation. One can see that irradiation
deforms the shape of the initial shock wave at $t>0$\,.
The region of the perturbed flow is wider at later times.
With increasing~$t$ the shock wave
becomes weaker (the density jump at the left hand side
diminishes). However, the absolute value of the shock
front velocity becomes larger as compared to the initial value~$D(0)$\,.

The acceleration of the shock front is clearly
visible in Fig.~\ref{figure1}. Indeed, the shift of the shock
front position from the point $x=0$ increases nonlinearly with
$t$\,. The reason for acceleration is rather simple:
as follows from \re{adsv} and Fig.~\ref{figure3} the sound
velocity $c_s$ behind the shock front increases with $t$
due to the heat deposition.

At given $t$ the density profile has three
characteristic points: 1)~the shock wave discontinuity at
$x=x_{\mathrm{sh}} (t)$\,, 2)~the intermediate kink
where the derivative $\ptx\rho$ jumps and 3)~the boundary of the
perturbed region on the right hand side. The position of
the third point can be found analytically: it is in fact
the characteristic
$x\raisebox{-.8ex}{$\scriptstyle C_+$}(t)$ with the initial
condition $x\raisebox{-.8ex}{$\scriptstyle C_+$}(0)=0$\,.
The calculation confirms that this point moves in
accordance with \re{char21} where ${\mathrm{v}}_0=v_{20}$ and
$c_0=c_{20}=\sqrt{\gamma P_{20}V_{20}}$\,.
It can be shown that the second point corresponds to
the characteristic $x\raisebox{-.8ex}{$\scriptstyle C_0$}(t)$ with
$x\raisebox{-.8ex}{$\scriptstyle C_0$}(0)=0$\,. The direct
numerical integration of \re{char13} gives the result
shown by the dotted curve in Fig.~\ref{figure1}. One can see that
on the $x-\rho$ plane the $C_0$ characteristic indeed goes
through the kink positions following from
Eqs. (\ref{fl31})--(\ref{fl33}).

\begin{figure}
\vspace*{-2cm}
\includegraphics[width=15cm]{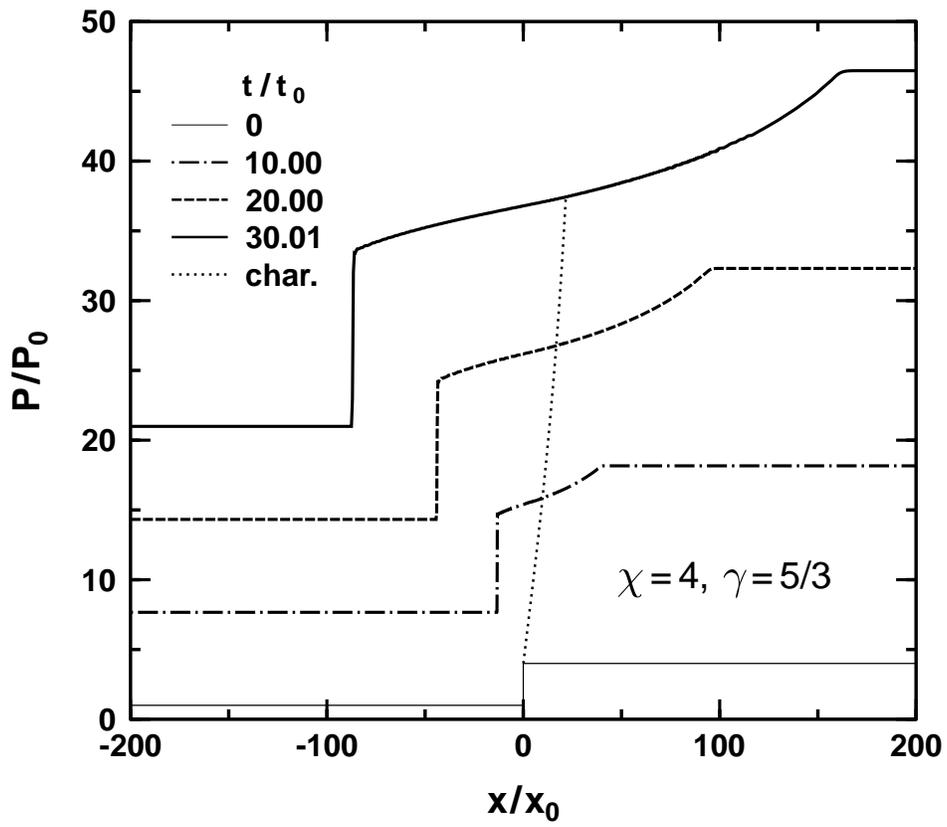}
\vspace*{-4.5cm}
\caption
{The same as in Fig.~\ref{figure1}, but for pressure profiles.}
\label{figure2}
\end{figure}

\begin{figure}
\vspace*{-2cm}
\includegraphics[width=15cm]{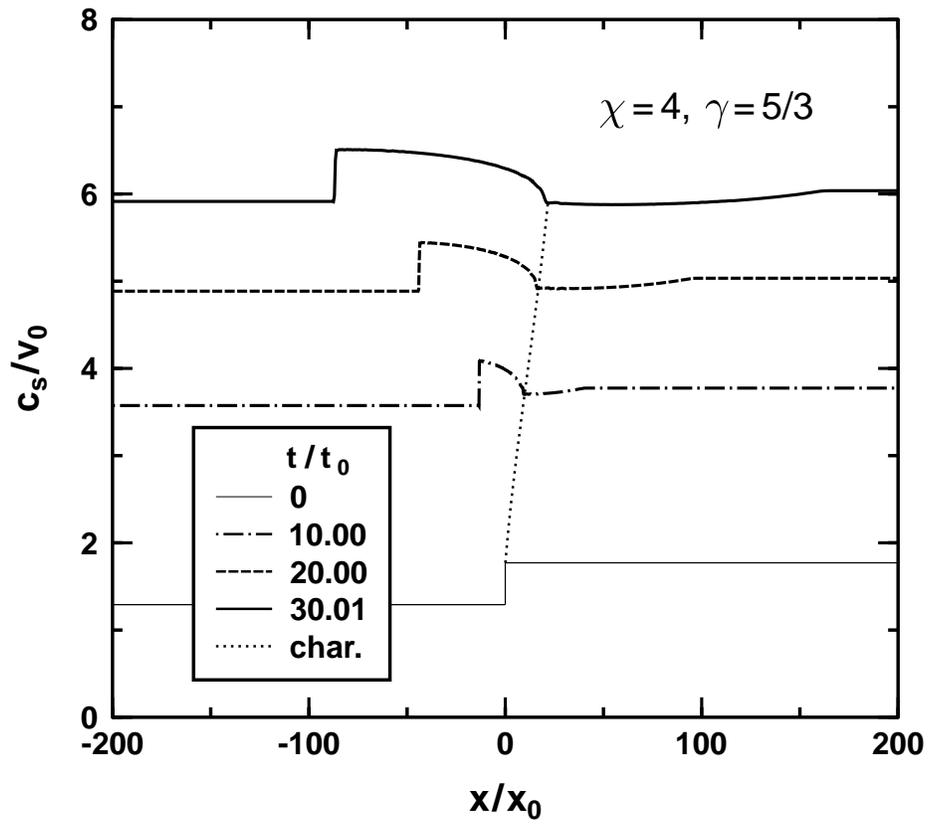}
\vspace*{-4.5cm}
\caption
{The same as in Fig.~\ref{figure1}, but for sound velocity profiles.}
\label{figure3}
\end{figure}

Profiles of pressure calculated for the same initial
conditions are shown in Fig.~\ref{figure2}. One can see that
the pressure jump at the shock front, $P_2-P_1$,
becomes larger at later stages of irradiation.
We have checked that the density and pressure
jumps at $x=x_{\mathrm{sh}}$
satisfy the Hugoniot adiabate given by \re{shwr}.
In accordance with \re{sf13}, pressure increases linearly
with $t$ at $x\to\pm\infty$\,. At known $\rho$ and $P$ the sound
velocity $c_s$ can be calculated by using the first equality
in~\re{adsv}. In the case of the ideal gas
$c_s$ is proportional to~$\sqrt{T}$\,. The results of
the calculation are shown in Fig.~\ref{figure3}. It is seen that
the jump in $c_s$, and therefore also in $T$\,, practically
does not increase with $t$\,. At large $t$ the asymptotic values
of $c_s$ at $x\to\pm\infty$ become nearly equal. This
agrees with \re{adsv}. Indeed, at $t\gg c^2_{20}/\lambda$ one
obtains the relations
\bel{assv}
\lim_{x\to\pm\infty}c_s\simeq\frac{3}{2}\hsp\xi_0\sqrt{\lambda t}\,,
\ee
where $\xi_0$ is defined by \re{xi0d}.

\begin{figure}
\vspace*{-2cm}
\includegraphics[width=15cm]{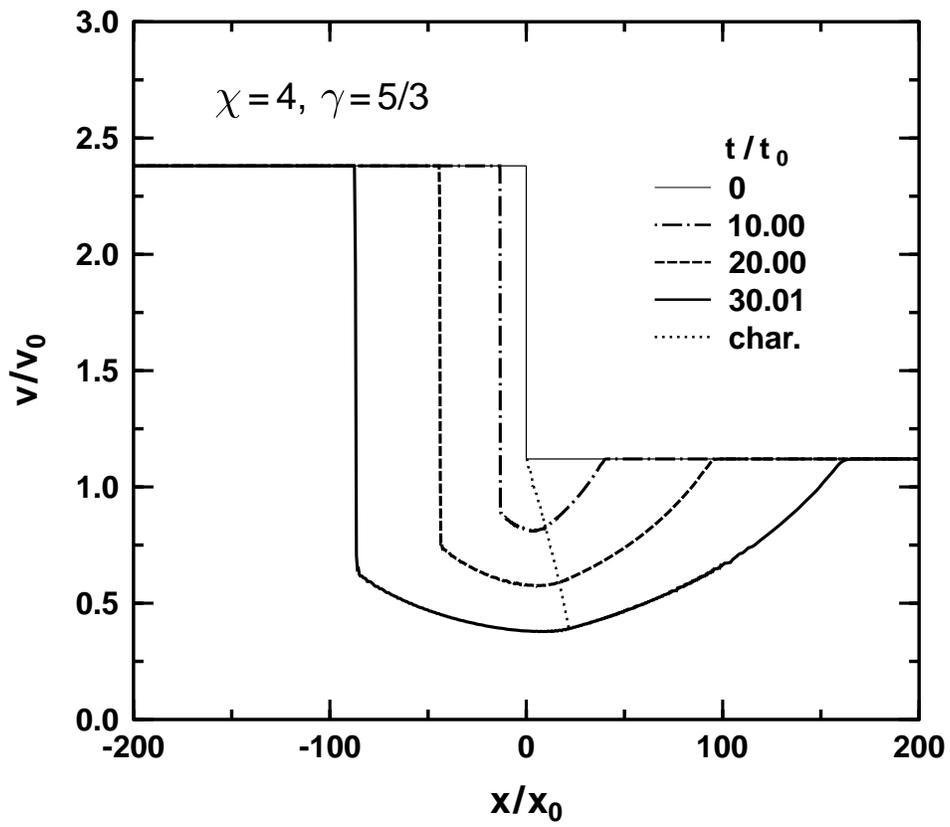}
\vspace*{-4cm}
\caption
{The same as in Fig.~\ref{figure1}, but for flow velocity profiles.}
\label{figure4}
\end{figure}

Fig.~\ref{figure4} shows the results for the flow velocity profiles.
Similar to density, values of $v$ do not change with $t$ at
$x\to\pm\infty$\,. However, the flow velocities in the
perturbed region diminish as compared to initial values.
As will be shown in the next section these velocities become
negative at late stages of irradiation.
Numerical values of the velocity jump agree with
conditions given in \re{shwr} (see footnote on page 7).
At large $t$ this jump increases approximately
proportional to $\sqrt{t}$\,.

\subsection{Self--similarity of flow at large irradiation
time}

The results presented in Sect.~IIIA suggest that at large
irradiation time, the fluid dynamical quantities exhibit
some properties of self--similarity. At $t\gg c_{20}^2/\lambda$
characteristic values of pressure and sound velocity increase,
respectively, like $t$ and $\sqrt{t}$\,. On the other hand,
at late stages of irradiation the size of the region with
a perturbed flow grows approximately like $t^{3/2}$\,.
This means that at \mbox{$t\to\infty$} the following
relations should hold asymptotically
\begin{eqnarray}
V&\simeq &V_{\mathrm{as}}=\rho_0^{-1}
\hat{V}(\xi)\,,\label{asr1}\\
v&\simeq&v_{\mathrm{as}}=
\sqrt{\lambda t}\,\hat{v}(\xi)\,,\label{asr2}\\
P&\simeq&P_{\mathrm{as}}=
(\gamma-1)\hsp\rho_0\lambda t\hsp\hat{P}(\xi)\,,\label{asr3}
\end{eqnarray}
where scaled dimensionless quantities
\mbox{$\hat{V},\,\hat{v},\,\hat{P}$} depends on $x$ and $t$
only via the self--similar variable
\bel{ssv}
\xi\equiv x/\sqrt{\lambda t^3}\,.
\ee
By substituting asymptotic expressions
(\ref{asr1})--(\ref{asr3}) into the equations
of fluid dynamics \mbox{(\ref{fl31})--(\ref{fl33})} one can
obtain the set of ordinary differential equations for
\mbox{$\hat{V},\,\hat{v}$} and $\hat{P}$ (see Appendix).
Apparently, these equations can not be solved analytically;
it is possible, however, to find the limiting
behaviour of flow directly from the numerical solution of
Eqs.~(\ref{fl31})--(\ref{fl33}) at large~$t$\,.

\begin{figure}
\vspace*{-2cm}
\includegraphics[width=15cm]{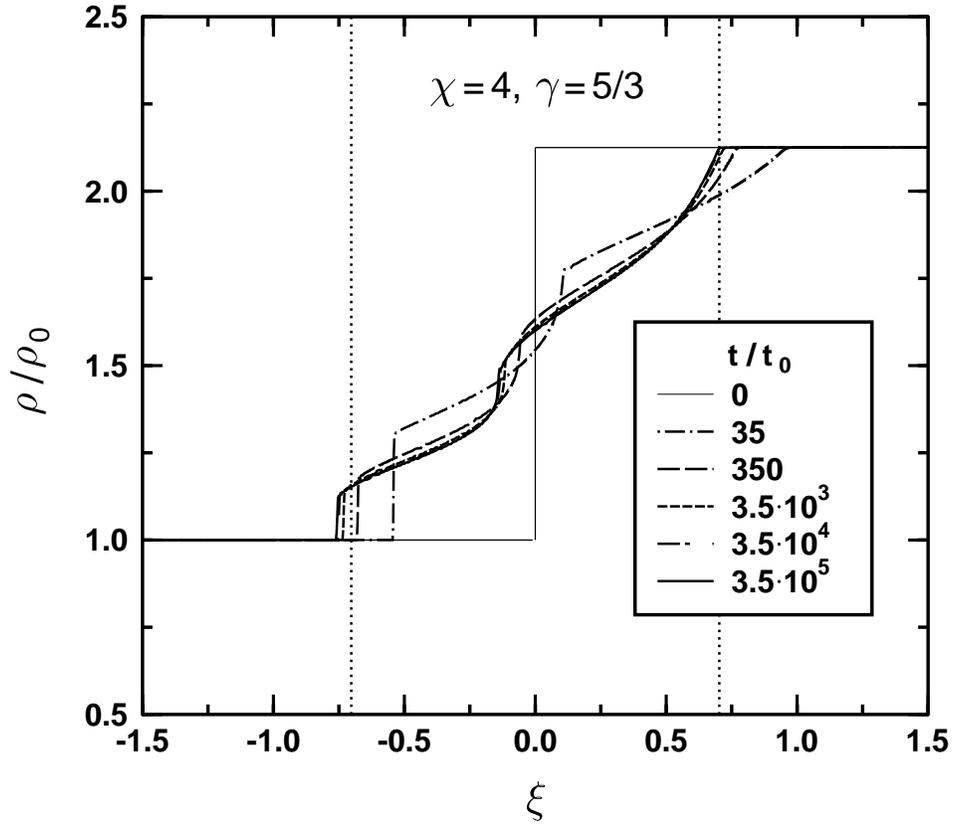}
\vspace*{-4cm}
\caption
{The density profiles at large irradiation times as
functions of \mbox{$\xi=x\,(\lambda\hsp t^3)^{-1/2}$}. The dotted
vertical lines show positions of the $C_\pm$
characteristics (\ref{asch}).}
\label{figure5}
\end{figure}

Fig.~\ref{figure5} represents density as function of $\xi$ at several
fixed times $t$\,. One can see that in agreement
with \re{asr1}, at large $t$ the density depends on $x$ and $t$
only via $\xi$\,.  The shock wave does not disappear as $t\to\infty$.
At late stages its front accelerates
in accordance with the relation
\bel{assw1}
x_{\mathrm{sh}}(t)\simeq \xi_{\mathrm{sh}}\sqrt{\lambda t^3}\,,
\ee
where $\xi_{\mathrm{sh}}$ is a constant fully determining
the asymptotic properties of the shock wave.
In particular, $\xi_{\mathrm{sh}}$ defines asymptotic behaviour of
the shock front velocity:
\nopagebreak
\bel{assw2}
D\hsp(t)\simeq -\frac{3}{2}\hsp\xi_{\mathrm{sh}}\sqrt{\lambda t}\,.
\ee
As seen in Figs.~5--6, the region of the perturbed (nontrivial)
flow corresponds to the interval $\xi_{\mathrm{sh}}<\xi<\xi_0$\,.
It will be shown below that the asymptotic position of the shock front
$\xi=\xi_{\mathrm{sh}}$ moves to the left with rising initial
pressure ratio $\chi$\,. For weak
initial shocks, $\chi\simeq 1$\,, the shock front becomes
close to the limiting position of the $C_-$
characteristics, i.e. \mbox{$\xi_{\mathrm{sh}}\simeq -\xi_0$}.

\begin{figure}
\vspace*{-2cm}
\includegraphics[width=15cm]{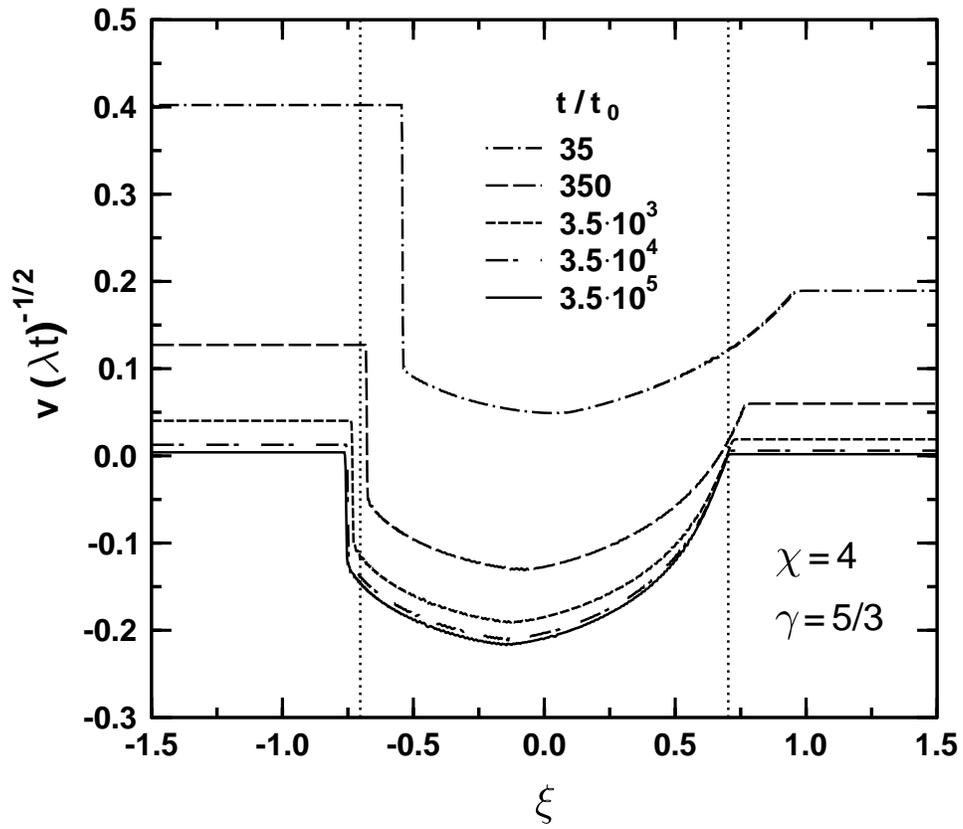}
\vspace*{-4cm}
\caption
{The same as in Fig.~\ref{figure5}, but for the scaled velocity profiles.}
\label{figure6}
\end{figure}

Another feature of the self--similar regime, is the appearance of a
rapid density variation at some intermediate $\xi_*$ behind the shock
front.  By analyzing Eqs.~(\ref{afl31})--(\ref{afl33}) (see Appendix)
one can show that at large $t$ the disturbed zone is characterized by
the combination of a strong discontinuity (the shock front at
\mbox{$\xi=\xi_{\mathrm{sh}}$}) and two weak discontinuities%
\footnote{%
 In such
 discontinuities spatial derivatives of fluid--dynamical
 variables are infinite or exhibit jumps.
}
at \mbox{$\xi=\xi_*$} and~$\xi_0$\,.  According to Fig.~\ref{figure6} the
convergence to the self--similar behaviour is much slower for the
velocity profiles.  It is interesting to note that the flow velocity
changes sign at late stages of irradiation.  Comparing Figs.~\ref{figure5}
and \ref{figure6}, one can see that the point $\xi=\xi_*$ corresponds to
the minimum of the asymptotic velocity profile $\hat{v}(\xi)$\,. This
point also gives the limiting position of the $C_0$ characteristics
at~$t\to\infty$\,. Indeed, substituting (\ref{asr2}) into \re{char13},
one can see that all $C_0$ characteristics satisfy the asymptotic relation
$x\raisebox{-.8ex}{$\scriptstyle C_0$}(t)\simeq\xi_*\sqrt{\lambda t^3}$\,,
where~$\xi_*$ is determined from the
equation%
\footnote{%
 According to \re{afl31}, $\hat{v}$ is minimal at $\xi=\xi_*$\,.
}
\bel{c0as}
\hat{v}\hsp(\xi_*)=\frac{3}{2}\hsp\xi_*\,.
\ee

\begin{figure}
\vspace*{-2cm}
\includegraphics[width=15cm]{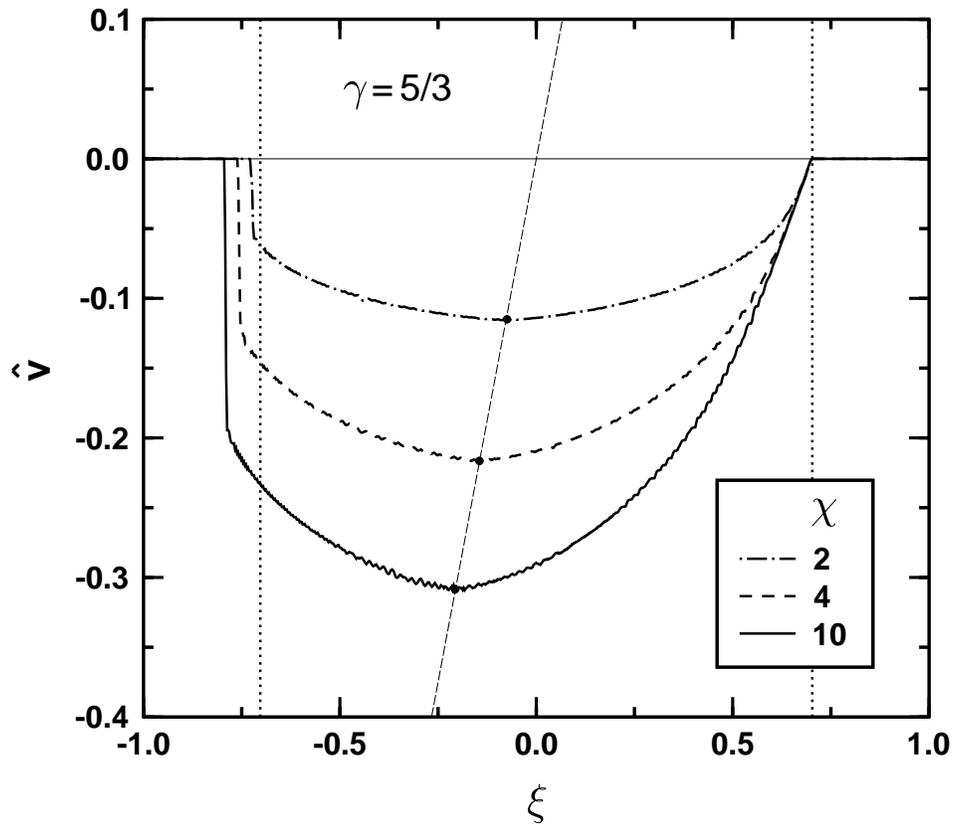}
\vspace*{-4cm}
\caption
{Asymptotic profiles of scaled velocity at various ratios of
initial pressure $\chi$\,. The thin dashed curve
corresponds to the line $3\hsp\xi/2$\,. The vertical dotted lines
indicate the points of
minimal velocities. Dotted lines are the same as in Fig.~\ref{figure5}.}
\label{figure7}
\end{figure}

\begin{figure}
\vspace*{-2cm}
\includegraphics[width=15cm]{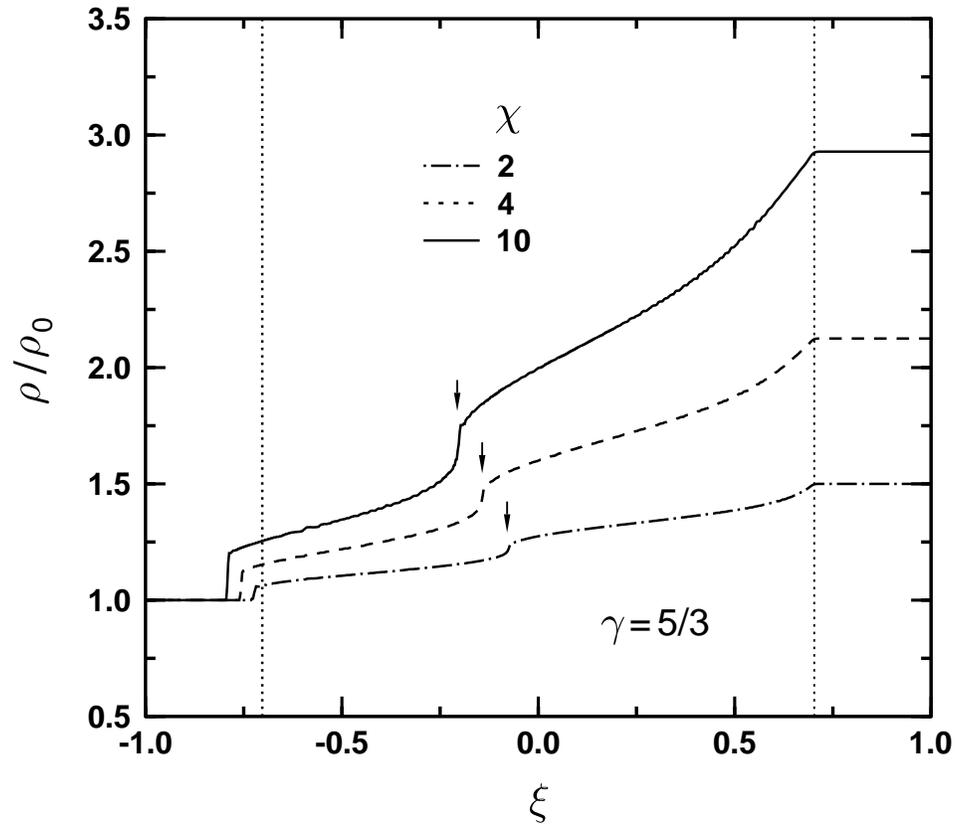}
\vspace*{-4cm}
\caption
{Asymptotic density profiles at various $\chi$\,.
The arrows indicate the values of $\xi_*$\,,
defined by \re{c0as}. The dotted lines are the same as in
Fig.~\ref{figure5}.}
\label{figure8}
\end{figure}

\begin{figure}
\vspace*{-2cm}
\includegraphics[width=15cm]{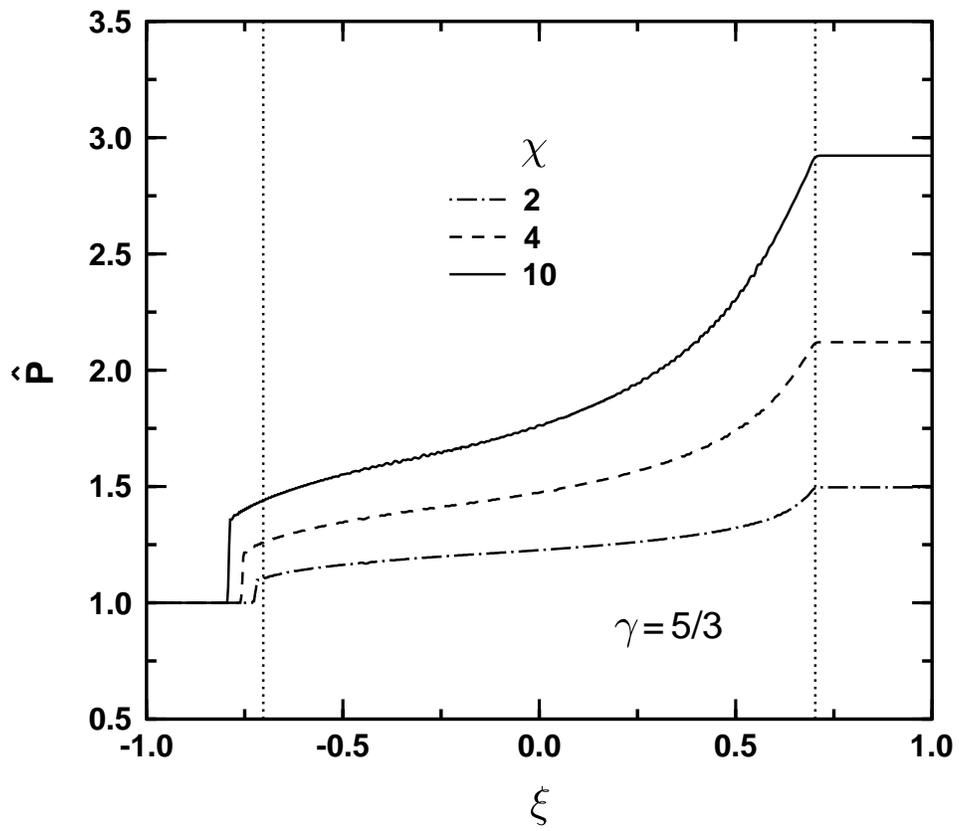}
\vspace*{-4cm}
\caption
{Asymptotic profiles of pressure at various $\chi$.
The dotted lines are the same as in Fig.~\ref{figure5}.}
\label{figure9}
\end{figure}

Figs.~7--9 show asymptotic profiles
of flow velocity, density and pressure calculated
for initial ratios $\chi=2,\,4$ and 10\,. In accordance with the above
discussion, the line $3\hsp\xi/2$ goes through minima of
the asymptotic velocity profiles at different $\chi$\,.
This is clearly seen in Fig.~\ref{figure7}. Asymptotic density profiles
are shown in Fig.~\ref{figure8}. Here the points $\xi_*$ are marked by
arrows. Fig.~\ref{figure9} shows the asymptotic pressure
profiles which, unlike the density profiles, vary smoothly at
$\xi=\xi_*$\,.

Jumps of asymptotic parameters at the shock front $\xi=\xi_{\mathrm{sh}}$
can be determined by using equations (\ref{shwr}) in the
limit \mbox{$t\to\infty$}. At $\xi<\xi_{\mathrm{sh}}$ one has $\hat{v}=0$
and $\hat{P}=\hat{V}=1$\,. Marking the flow parameters
at $\xi=\xi_{\mathrm{sh}}+0$ by indices 'sh', one gets the relations
\bel{shwr1}
-\hsp\frac{3}{2}\,\xi_{\mathrm{sh}}=
\frac{\ds\hat{v}_{\mathrm{sh}}-\frac{3}{2}\,\xi_{\mathrm{sh}}}
{\hat{V}_{\mathrm{sh}}}=\sqrt{(\gamma -1)\hsp\frac
{\hat{P}_{\mathrm{sh}}-1}{1-\hat{V}_{\mathrm{sh}}}}=\frac{3}{2}\,\xi_0\,
\sqrt{\frac{1+\hat{P}_{\mathrm{sh}}}{1+\hat{V}_{\mathrm{sh}}}}\,.
\ee
Using these conditions,
$\hat{v}_{\mathrm{sh}}\,, \hat{P}_{\mathrm{sh}}$ and $\hat{V}_{\mathrm{sh}}$
may be obtained
as functions of $\xi_{\mathrm{sh}}$\,. For example, the scaled pressure
behind the shock front is equal to
\bel{sprsw}
\hat{P}_{\mathrm{sh}}=1+\frac{2\hsp\gamma}
{\gamma +1}\left(\frac{\xi_{\mathrm{sh}}^2}{\xi_0^2}-1\right).
\ee
Substituting in \re{shwr1} the values of $\xi_{\mathrm{sh}}$ determined from
numerical solution of fluid--dynamical equations,
one may calculate asymptotic parameters of the shock wave
at given~$\chi$\,.
\begin{table}[t]
\caption{Parameters of asymptotic flow behind the shock wave
($\gamma=5/3$)}
\vspace*{3mm}
\label{tab1}
\begin{center}
\begin{tabular}{|c|c|c|c|c|c|} \hline\hline
$\chi$&$\xi_{\mathrm{sh}}$&$\hat{V}_{\mathrm{sh}}^{-1}$&
$\hat{v}_{\mathrm{sh}}$&~$\hat{P}_{\mathrm{sh}}$&$\xi_*$\\
\hline
~2~& -0.720 & 1.038 & -0.040 & 1.062 & -0.074\\
~4~& -0.759 & 1.120 & -0.122 & 1.208 & -0.144\\
~10~& -0.785 & 1.176 & -0.175 & 1.310 & -0.209\\
\hline\hline
\end{tabular}
\end{center}
\end{table}
\noindent
The results of this calculation are shown in Table~\ref{tab1}.
It has been checked that these results
agree well with jumps of flow velocity, density and pressure
in Figs.~7--9. The last column gives the points where
asymptotic velocities achieve their minima.

\section{Simulation of irradiated shocks under laboratory
conditions}

The investigation of the effects studied in this paper in the
laboratory will be possible by depositing energy into a target with a
long deposition length. The only practical means for doing this is
the use of heavy--ion beams. We have therefore performed realistic
hydrodynamic simulations to estimate the magnitude of the effect
with heavy--ion beams such as will be available with the proposed
enhanced accelerator \cite{Gsi01} at the GSI laboratory, Germany.
In all calculations presented in this section we consider
cylindrical targets made of solid Gold. The targets were irradiated
by singly ionized Uranium beams collinear with target axes.
In all cases the bombarding energy of heavy ions was 500 MeV per
nucleon.

The simulations were performed using the code \textsl{CAVEAT} developed
at Los Alamos~\cite{Add92}. The code solves the fluid--dynamical
equations using the \textit{arbitrary Lagrangian--Eulerian}
(\textsl{ALE}) technique and the second--order Godunov--type scheme.
Local values of ion energy deposition were calculated by decomposing
the beam into a large number of beamlets and tracing them through the
moving cells.  Effects of thermal conductivity and viscosity were
neglected since they are not important for the considered problem.

In the first simulation the target had 3~mm radius and 6~mm
length. It was assumed that the beam had constant radial and temporal
profiles with the beam radius $r_b=2$ mm and the pulse duration
$\tau=$ 50~ns. The total penetration range of a 500 MeV/A Uranium ion
in solid Gold of normal density is about 3.5~mm.  The total deposited
energy of the beam was 38.1~kJ.  In the considered case the specific
deposition rate $\lambda(x)$ can be evaluated as
\bel{endri}
\lambda(x)=\frac{N}{\pi r_b^2\tau}\frac{dE}{\rho dx}\,.
\ee
Here $dE/dx$ is the energy loss per unit path length of a single ion in
a target material of the density $\rho$\,, $N$ is the total number of
bombarding ions. The ion deposition profiles $dE/dx$ have been
calculated by using a realistic deposition model realized in the
\textsl{TRIM} code~\cite{Zie85}. The equation of state for Gold was
taken from the \textsl{SESAME}~\cite{Lyo92} tables. In fact, we solved
the fluid--dynamical equations analogous to
Eqs.~(\ref{fl21})--(\ref{fl23}), but generalized for the case of
two--dimensional (axially symmetric) cylindrical flow.

\begin{figure}
\vspace*{-2cm}
\includegraphics[width=15cm]{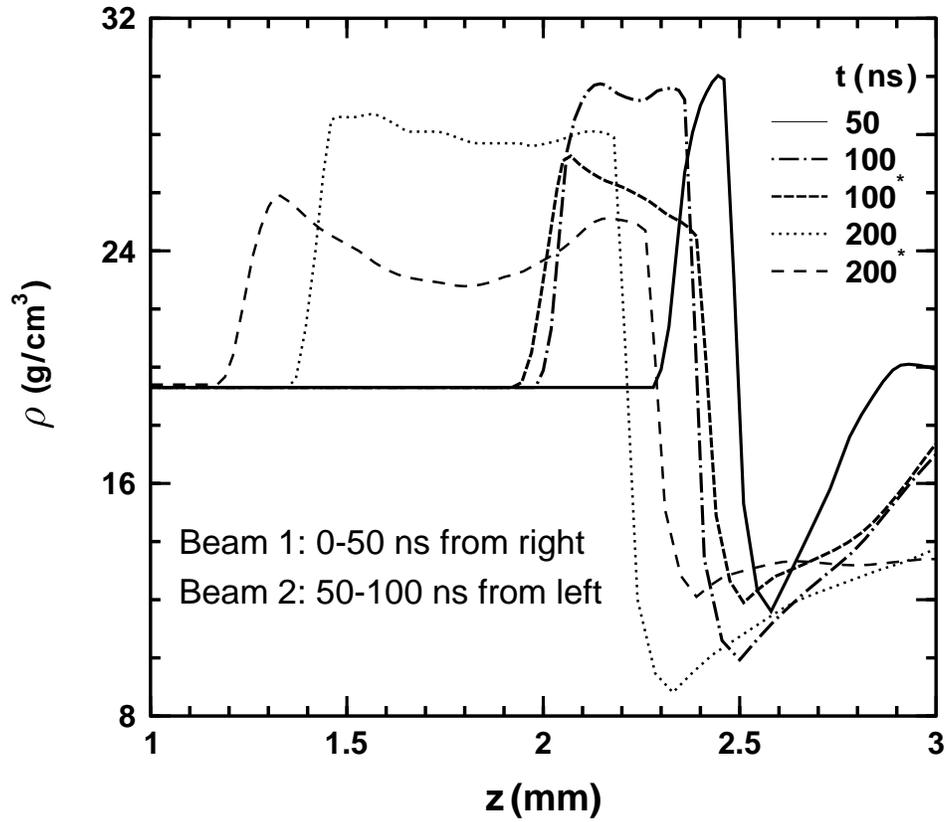}
\vspace*{-4cm}
\caption
{Profiles of density along the axis of a cylindrical solid Gold
target irradiated by a heavy--ion beam of 50~ns duration from the
right at various times (the parameters of the beam are
given in the text). The curves for times marked with an asterisk
correspond to the situation with an additional beam of the same
properties impinging on the target from the left after the first beam
has ended.}
\label{figure10}
\end{figure}

Two situations were compared in our simulations: irradiation with one
beam from the right, and the same with a second beam of identical
properties from the left starting with a 50 ns delay, i.e. exactly
after the end of the first beam. The target length is chosen such that
the second beam completely penetrates the running shock wave, and
although the realistic deposition profile is used, the second Bragg
peak is sufficiently far to the right of the shock. As a consequence,
the local deposition from the second beam is approximately constant
inside the shock. Fig.~\ref{figure10} compares density profiles along the
target axis ($r=0$) for the two cases at several times between 50 and
200~ns. At $t=$ 50~ns the density profile is, of course, not yet
disturbed by the second beam. One can clearly see that the shock
without irradiation runs towards the left, losing speed and amplitude
rather slowly at $t$ between 50 and 200 ns.  A weaker shock caused by
the backward side of the Bragg peak is not visible since its position
is to the right of the region plotted in Fig.~\ref{figure10}.  The decay in the
shock strength due to the additional deposition of energy is clearly
visible already at $t=$ 100 ns. The shock moves faster and the density
of matter behind its front diminishes in accordance with the
predictions of the simplified model studied in Sect.~III. At
\mbox{$t=$ 200~ns} the density jump is reduced by about 30--50\%.
The shock speed has increased by about 20\%, which should be clearly
observable in experiments. The additional structure in the compressed
region is apparently caused by the non--uniform deposition of the beam.

\begin{figure}
\includegraphics[width=15cm]{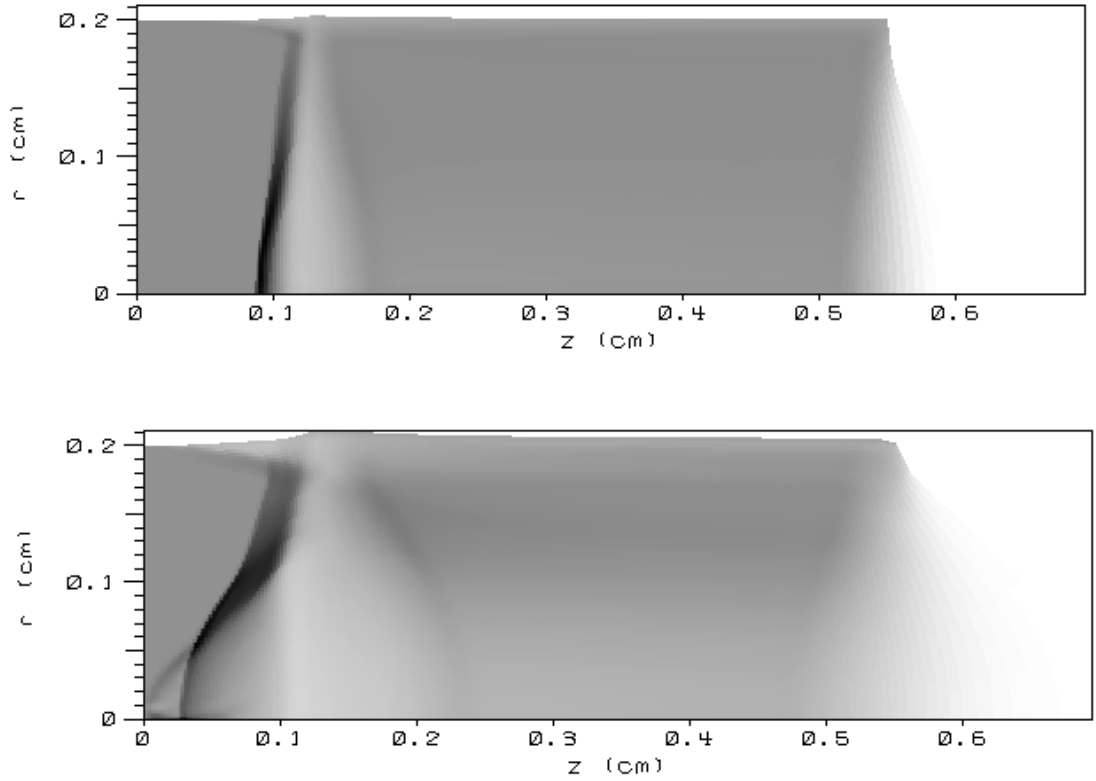}
\vspace*{3cm}
\caption
{Density plots of a Gold cylinder irradiated by a heavy--ion
beam from the right. The darkness of shading is
proportional to the density of matter. Upper and lower plots correspond
to irradiation times 6 and 12~ns, respectively. The curved shock front is
clearly developed after 6~ns, but is destroyed in the center already
at 12~ns. The effects of beam penetration through the shock
front are also visible at this time. Note that the grey scale
is different in two plots: it was chosen to show
geometrical effects more clearly.}
\label{figure11}
\end{figure}

Another possible scenario may exist even for a single beam irradiation.
In the case when the beam has long--enough duration, the sideward
expansion of the target material will result in depletion of density
and, therefore, in a longer range of bombarding particles near the
target axis. This enhanced range can lead to beam penetrating through
the shock wave created at the initial stages of target irradiation. As
a consequence, at late stages the shock wave will be weaker in a
central region of the target. An example of such configuration is
shown in Fig.~\ref{figure11}. In this case the target had a radius of 2~mm
and a length of 5~mm. The beam had the same properties as above, except
that its radial profile was assumed to be Gaussian with a full width at
half maximum of 1.5~mm. The Gaussian profile was chosen not only
because it is more realistic, but also because in this case the density
starts decreasing immediately after beginning of irradiation even near
the axis of the target. The results of calculation demonstrate how the
original curved shock front created by the beam is destroyed in the
center, leading to a ring--like structure. This effect can be observed
after the shock wave emerges from the other end of the target.

To check whether the effect might be observable with heavy--ion beams
which will be available at the first stage~\cite{Gsi01} of the upgraded
GSI facility we have performed two additional calculations for smaller
beam intensities. In the first calculation both beams (the first one
generating the shock and the second beam irradiating it) had equal
total energies, reduced by factor 5 as compared to previous cases.
We chose the beam duration \mbox{$\tau=$}~300~ns and radius
\mbox{$r_b=$} 5~mm. The target had 3~mm radius and 6~mm length. In the
second calculation the beam generating the shock wave was replaced by a
stronger beam with parameters considered in the first series of
simulations (see Fig.~\ref{figure10} and footnote on page 2).  In both
calculations the modifications of shock velocities and compression
ratios turned out to be at the percentage scale which is below the
measurable level. Therefore, experimental investigation of the effects
predicted in this paper requires intensive heavy--ion beams which will
be available only at the fully upgraded GSI facility.

\section{Summary and outlook}

In this paper we have investigated the behaviour of a shock wave
stimulated by energy deposition. The case of a planar shock wave
under the influence of homogeneous irradiation has been studied in
detail. This was done by solving the one--dimensional
fluid--dynamical equations. It is shown that the initial shock
discontinuity decays into accelerated shock wave and a simple wave
propagating into the post--shock matter. It is found that the flow
exhibits a self--similar behaviour at late times.

The fluid--dynamical simulations have also been performed
for the more realistic conditions expected for laboratory
experiments. We have studied the behaviour of a shock wave
under the influence of additional energy deposition
produced by a heavy--ion beam in a cylindrical target
and found that the main features of the above planar
solution are also reproduced in this more realistic situation.
Another interesting effect studied is the weakening of the shock
wave in the vicinity of the target axis due to the sideward
expansion of matter and the increased deposition length of
bombarding ions. These examples
represent possible experimental set--ups for studying
the effects of energy deposition on the structure of shock
waves. Such experiments are possible at any laboratory
having intensive heavy--ion beams with high enough
deposition energies.

In the future we are planning to study analogous
processes in the case of spherical geometry and nonuniform
density distributions that is more relevant for supernovae
explosions. Another problem is to study the
influence of cooling and internal heat transport
on the shock wave dynamics. In particular, it is interesting
to investigate the role played by endo-- and exothermic
nuclear reactions during the propagation of shock front.

\section*{Acknowledgments}

This work has been supported by the RFBR Grant No.~00--15--96590,
GSI, BMBF and WTZ.

\section*{Appendix: Equations for asymptotic flow profiles}
\setcounter{equation}{0}
\renewcommand{\theequation}{A\arabic{equation}}

Inserting expressions (\ref{asr1})--(\ref{asr3})
into Eqs.~(\ref{fl31})--(\ref{fl33}) one obtains the
coupled set of ordinary differential equations
for asymptotic profiles \mbox{$\hat{V}(\xi),\,\hat{v}(\xi)$} and
$\hat{P}(\xi)$\,:
\begin{eqnarray}
&&\left(\hat{v}-\frac{3}{2}\hsp\xi\right)\hat{V}^{\hsp\prime}
=\hat{v}^{\hsp\prime}\hsp\hat{V}\,,\label{afl31}\\
&&\left(\hat{v}-\frac{3}{2}\hsp\xi\right)
\hat{v}^{\hsp\prime}+(\gamma-1)\hsp\hat{P}^{\hsp\prime}\hsp\hat{V}
=-\frac{\hat{v}}{2}\,,\label{afl32}\\
&&\left(\hat{v}-\frac{3}{2}\hsp\xi\right)
\hat{P}^{\hsp\prime}+(\gamma\hat{v}^{\hsp\prime}+1)\hsp\hat{P}
=\frac{1}{\hat{V}}\,,\label{afl33}
\end{eqnarray}
where primes denote derivatives with respect to $\xi$\,.
The boundary conditions for these equations may be derived
from Eqs.~(\ref{sf11})--(\ref{sf13}). At $t\to\infty$\,, using
(\ref{asr1})--(\ref{asr3}) one has
\begin{eqnarray}
\hat{v}\hsp(\pm\infty)&=&0\,,\label{bc1}\\
\hat{P}\hsp(-\infty)&=&\hat{V}\hsp(-\infty)=1\,,\label{bc2}\\
\hat{P}\hsp(+\infty)&=&\hat{V}^{-1}\hsp(+\infty)=
\rho_{20}/\rho_0\,.\label{bc3}
\end{eqnarray}
The above equations are trivially satisfied in the regions
$\xi>\xi_0$ and $\xi<\xi_{\mathrm{sh}}$\,. Therefore it is
sufficient to solve (\ref{afl31})--(\ref{afl33}) only in the
interval $(\xi_{\mathrm{sh}},\hsp\xi_0)$\,. The boundary
conditions at $\xi=\xi_0$ are given by Eqs.~(\ref{bc1}),
(\ref{bc3}) with the replacement $+\infty\to\xi_0$\,.
The parameters of asymptotic flow at $\xi=\xi_{\mathrm{sh}}+0$
can be determined from \re{shwr1}. They are functions
of the shock wave position $\xi_{\mathrm{sh}}$\,.

One can rewrite Eqs.~(\ref{afl31})--(\ref{afl33}) in the
vector form
\bel{aflv}
A\hat{\bm{y}}^{\hsp\prime}=\bm{B}\,,
\ee
where $\hat{\bm{y}}$ is a column with the components
\mbox{$\hat{V},\hsp\hat{v},\hsp\hat{P}$}\,. The matrix
$A$ and the vector $\bm{B}$ may be easily obtained from the
above set of equations. The derivatives
$\hat{\bm{y}}^{\hsp\prime}$ and the matrix $A^{-1}$
do not exist at the points $\xi$ where
\bel{adet}
{\mathrm{det}}\hsp A=\left(\hat{v}-
\frac{3}{2}\hsp\xi\right)\left[\left(\hat{v}-
\frac{3}{2}\hsp\xi\right)^2-\hat{c}^2\right]=0\,.
\ee
Here \mbox{$\hat{c}=\frac{\ds 3}{\ds
2}\hsp\xi_0\sqrt{\hat{P}\hat{V}}$}
is the scaled sound velocity. Comparing this equation with
Eqs.~\mbox{(\ref{char11})--(\ref{char13})} one can see that solutions of
(\ref{adet}) give the asymptotic positions of the $C_\pm$ and
$C_0$ characteristics. The latter is defined by \re{c0as}.

In fact, the right boundary of the nontrivial region, $\xi=\xi_0$\,,
is the $C_+$ characteristic. At this point
$\hat{v}^{\hsp\prime}$ jumps from zero (at $\xi>\xi_0$) to
some finite value $a_v\equiv\hat{v}^{\hsp\prime}(\xi_0-0)$\,.
By using Eqs.~(\ref{afl31})--(\ref{afl33}) for small
$\xi_0-\xi$\,, one can show that their solution between
the characteristics $C_+$ and $C_0$\,, i.e. in the interval
$(\xi_*,\hsp\xi_0)$\,, is a functional of the parameters $a_v$
and~$\rho_{20}/\rho_0$\,.
At given $a_v$ one can solve (\ref{afl31})--(\ref{afl33})
in the whole interval. Below this solution is marked by
the subscript ``+''.

On the other hand, choosing some value of $\xi_{\mathrm{sh}}$
and calculating $\hat{\bm{y}}\hsp(\xi_{\mathrm{sh}}+0)$\,,
one can find the solution of Eqs.~(\ref{afl31})--(\ref{afl33})
between the shock front and the $C_0$ characteristic,
i.e. in the interval $(\xi_{\mathrm{sh}},\hsp\xi_*)$\,.
The corresponding solution (denoted below by the subscript
``--'') is fully determined by the parameter $\xi_{\mathrm{sh}}$\,.
Matching the solutions $\hat{\bm{y}}_+$ and $\hat{\bm{y}}_-$ at
$\xi=\xi_*$ gives three conditions for determining three unknown
parameters $\xi_{\mathrm{sh}},\,\xi_*$ and $a_v$:
\begin{eqnarray}
\hat{v}_-(\xi_*)&=&\hat{v}_+(\xi_*)=
\frac{3}{2}\hsp\xi_*\,,\label{mc11}\\
\hat{P}_-(\xi_*)&=&\hat{P}_+(\xi_*)\,.\label{mc12}
\end{eqnarray}
By using Eqs.~(\ref{mc11})--(\ref{mc12}) and
(\ref{afl31})--(\ref{afl33}) it may be shown that $\hat{V}$
will be also continuous at~$\xi=\xi_*$\,.

A ``shooting'' method may then be used to find
$\xi_{\mathrm{sh}}$ and asymptotic flow profiles
$\hat{\bm{y}}(\xi)$ at different pressure ratios $\chi$\,.
These calculations agree well with the results
obtained in Sect.~IIIB.

%
%
%

\end{document}